\documentstyle[11pt,aaspp4,flushrt]{article}

\def\chandra{{\it Chandra}} 
\def\asca{{\it ASCA}} 
\def\rosat{{\it ROSAT}} 
\def\lum{erg s$^{-1}$}
\def\flux{erg cm$^{-2}$ s$^{-1}$}
\def\nh{cm$^{-2}$}

\begin{document}

\title{\chandra~ uncovers a hidden Low-Luminosity AGN \\ 
in the radio galaxy Hydra~A (3C~218)}

\author{Rita M. Sambruna, George Chartas, and Michael Eracleous}
\affil{The Pennsylvania State University, Department of Astronomy and
Astrophysics, 525 Davey Lab, State College, PA 16802 (email:
rms@astro.psu.edu)} 

\author{Richard F. Mushotzky}
\affil{NASA/GSFC, Code 662, Greenbelt, MD 20771}

\author{John A. Nousek} 
\affil{The Pennsylvania State University, Department of Astronomy and
Astrophysics, 525 Davey Lab, State College, PA 16802}

\begin{abstract}

We report the detection with \chandra ~of a Low-Luminosity AGN (LLAGN)
in the Low Ionization Emission Line Region (LINER) hosted by Hydra~A,
a nearby ($z$=0.0537) powerful FR~I radio galaxy with complex radio
and optical morphology. In a 20 ks ACIS-S exposure during the
calibration phase of the instrument, a point source is detected at
energies $\gtrsim$ 2 keV at the position of the compact radio core,
embedded in diffuse thermal X-ray emission ($kT \sim 1$ keV) at softer
energies. The spectrum of the point source is well fitted by a heavily
absorbed power law with intrinsic column density N$_H^{int} \sim 3
\times 10^{22}$ \nh~ and photon index $\Gamma \sim 1.7$. The intrinsic
(absorption-corrected) luminosity is $L_{2-10~keV} \sim 1.3 \times
10^{42}$ \lum. These results provide strong evidence that an obscured
AGN is present in the nuclear region of Hydra~A. We infer that the
optical/UV emission of the AGN is mostly hidden by the heavy intrinsic
reddening. In order to balance the photon budget of the nebula, we
must either postulate that the ionizing spectrum includes a UV bump or
invoke and additional power source (shocks in the cooling flow or
interaction with the radio jets).  Using an indirect estimate of the
black hole mass and the X-ray luminosity, we infer that the accretion
rate is low, suggesting that the accretion flow is advection
dominated. 
Finally, our results support current unification schemes for
radio-loud sources, in particular the presence of the putative
molecular torus in FR~Is.  These observations underscore the power of
the X-rays and of \chandra~in the quest for black holes.

\end{abstract} 

\noindent {\underline{\em Subject Headings:}} Galaxies: active --- 
galaxies: individual (Hydra A, 3C~218) --- X-rays: galaxies. 

\section{Introduction} 

Recent {\it HST} and ground-based optical observations provided strong
evidence that many nearby galaxies harbor supermassive black holes
(e.g., Kormendy \& Richstone 1995), possibly accreting at
sub-Eddington luminosities (Fabian \& Rees 1995). About 40\% of nearby
early-type galaxies exhibit signs of mild nuclear activity, in the
form of weak non-thermal radio cores (Sadler et al. 1989) and Low
Ionization Emission Lines (LINERs; Heckman 1986; Ho, Fillippenko, \&
Sargent 1997a). These results collectively suggest that many nearby
galaxies may harbor weak nuclear activity in the form of a
Low-Luminosity Active Galactic Nucleus (LLAGN; e.g., Ho 1999a).

Thanks to their high penetrating power, X-rays provide an optimal
window to search for weak nuclear activity.  Indeed, previous X-ray
\rosat~ images of a handful of galaxies show the presence of a central
unresolved nucleus within the 5\arcsec~ HRI resolution (e.g., Fabbiano
1996). Indirect clues are provided by \asca~ spectral constraints: a
heavily absorbed power law component is often measured at energies
$\gtrsim 2$ keV, with intrinsic luminosities L$_{2-10~keV} \sim
10^{40-42}$ \lum~, photon indices $\Gamma_{2-10~keV} \sim 1.5-1.7$,
and a narrow Fe emission line at 6--7 keV in a few cases, suggestive
of a LLAGN (Makishima et al. 1994; Ptak et al. 1999; Sambruna,
Eracleous, \& Mushotzky 1999; Terashima et al. 1999). Within the
coarse angular resolution of these detectors, alternative scenarios
can not be ruled out in many cases (e.g., a starburst or X-ray
binaries). Unambiguous evidence for nuclear activity would be provided
by the detection of a point source at the galaxy center.  With its
unprecedented angular resolution (0.5\arcsec), wide-band coverage
(0.2--10 keV), and high sensitivity, \chandra~ is uniquely suited to
this task. 

In this Letter, using recent \chandra~calibration observations we
discover a LLAGN in the LINER harbored by the nearby cD galaxy Hydra~A
($z$=0.0537).  Host of the powerful FR~I radio source 3C~218, Hydra~A
is the dominant member of the poor cluster of galaxies A780, and
famous for its twin-jet radio morphology (Taylor et al. 1990).
The nuclear region contains a $\sim$ 6\arcsec~emission-line nebula
(Baum et al. 1989; Heckman et al. 1989) and a disk of star formation
(McNamara 1995; Melnick, Gopal-Krishna, \& Terlevich 1997). At the
position of the nucleus, a LINER-like optical and UV spectrum is
observed (Hansen, J{\o}rgensen, \& N{\o}rgaard-Nielsen 1995), which,
together with the powerful lobe radio emission, led to the
classification of Hydra~A as a Weak Line Radio Galaxy (Tadhunter et
al. 1998). The X-ray emission from Hydra~A in previous {\it Einstein},
\rosat, and \asca~ images is dominated by the cluster and the inner
($\lesssim$ 1.5\arcmin) cooling flow, with $L^{cluster}_{0.5-4.5~keV}
\sim 2 \times 10^{44}$ \lum~ (David et al. 1990; Ikebe et al. 1997).

In the following, we assume a Friedman cosmology with $H_0=75$ km
s$^{-1}$ Mpc$^{-1}$ and $q_0=0.5$. At the luminosity distance of
Hydra~A (217.46 Mpc), 1\arcsec=0.95 kpc.

\section{Observations and Data analysis} 

Hydra~A was observed with \chandra~ ACIS-S for 20 ks on 1999 November
11 at the aimpoint of S3 in faint telemetry mode with 5 CCDs turned
on. 
The data were analyzed using the \verb+EventBrowser+ tool at Penn
State, and the \verb+CIAO+ software provided by the \chandra~X-ray
Center. Only events for \asca~ grades 0, 2, 3, 4, and 6 were
accepted. The S3 background was rather stable during the observation.

Spectral analysis was performed within \verb+XSPEC+ v.10.0 using
response files appropriate for the S3 aimpoint and for the epoch of
the present observation (\verb+ccd7_c0.7.15.32.rmf+,
\verb+s3_c1_middle.arf+). The spectra were rebinned over the energy
range 0.2--6 keV to have a minimum of 20 counts in each bin, to
validate the use of the $\chi^2$ statistics.  With a total count rate
of $\sim$ 0.02 c/s, pileup is not a concern. At this time of writing,
there appears to be an uncertainty of the order of 10\% in the total
effective area of the telescope mirror in the 1--2 keV energy
range. Fortunately, for the present analysis the energy range of
interest for the nuclear properties is restricted to energies above 2
keV, where the AGN component dominates.  The errors in the on-axis
effective area above 2 keV are of the order of 5\% (Jerius et
al. 2000). 

Hydra~A was also observed with ACIS-I for 20 ks in 1999 October 30,
after the CCDs suffered radiation damage 
resulting in a degradation of the gain and spectral resolution. 
To improve the signal-to-noise ratio we extracted the ACIS-I spectrum
of the nucleus in a region of similar size as for the ACIS-S
data, and analyzed the data using a position-dependent 
spectral response including an empirical CTI correction.  The 2--6 keV
ACIS-I spectrum of the nucleus was fitted simultaneously to the ACIS-S
data leaving the intercalibration factors free to vary, giving a ratio
of the two normalizations of $N_{ACIS-S}/N_{ACIS-I}=1.09 \pm 0.22$.

\section{Results} 


Figure 1 shows the S3 image of the central regions of Hydra~A in
0.2--10 keV, with the VLA radio contours overlayed (Taylor et
al. 1990).  The ACIS-S data were adaptively smoothed using a Gaussian
function with a kernel of 3.5\arcsec~, and shifted by 3.6\arcsec~to
align the X-ray and radio core and the other X-ray/radio morphology to
better than 0.5\arcsec. This shift is well within the range of
uncertainties on the astrometry (0.5\arcsec -- 5\arcsec) measured in
several other ACIS observations during the orbital activation and
check-out phases.  X-ray emission from the compact radio core is
readily apparent, embedded in diffuse soft X-ray emission. In the
0.2--10 keV band, $\sim$ 900 counts are detected within
2.5\arcsec~from the position of the radio core and the radial profile
of the X-ray source is extended.
At energies $\gtrsim$ 2 keV, the radial profile is point-like and a
total of $\sim$ 150 counts are collected in 2--10 keV within
1.5\arcsec~ (or 80\% of the total encircled energy). The hard X-ray
point source is also present in the 20 ks ACIS-I exposure. Thus, with
\chandra~we were able to detect a point-like X-ray source for the
first time in Hydra~A at hard energies, indicating that the core radio
emission is due to a hidden AGN.  Interestingly, no X-ray emission is
detected from either the jets or lobes (Fig. 1). On the contrary, the
extended radio structures appear to occupy regions relatively
deficient in X-ray photons. This is opposite to what recently detected
with \chandra~ in the FR~II radio galaxy 3C~295 (Harris et al. 2000).


Figure 2 shows the inner 10\arcsec~ region around the nucleus in
contour form, in 3.7--4.3 keV.  The nuclear point source is apparent,
together with extended faint emission elongated by $\sim$ 3\arcsec~ in
a N-W direction. Note the extension in the same direction in the radio
contours (Fig. 1).  At roughly this distance, a spot of enhanced blue
emission was observed in previous $B$ and $U$ band images, identified
as the bluest edge of a star forming disk (McNamara 1995; Melnick et
al. 1997). A total of $\sim$ 260 counts are detected within 2\arcsec~
over the 0.2--10 keV band for this region.  The study of the starburst
and of the large-scale X-ray emission in Hydra~A are beyond the scope
of this paper, and will not be discussed any further (see McNamara et
al. 2000).
 

The nuclear X-ray spectrum was extracted from a circular region with
radius 1.5\arcsec. The background was extracted in an annulus centered
on the nucleus with inner and outer radii 1\arcmin~ and 1.3\arcmin,
respectively. The cluster contribution in an area similar to the
nuclear region is small, $\sim$ 6\%, while the instrumental and cosmic
background is negligible, $\sim 2 \times 10^{-5}$ ph s$^{-1}$
arcsec$^{-2}$.  The spectrum was fitted with a two-component model,
including an absorbed power law at energies $\gtrsim$ 2 keV, and a
Raymond-Smith thermal plasma at softer energies. This model is a very
good description of the ACIS-S data, with $\chi^2$=17 for 23 degrees
of freedom. The data and residuals are shown in Figure 3.  All
spectral components in the fit were absorbed by the Galactic column
density, N$_H^{Gal}=4.9 \times 10^{20}$ cm$^{-2}$ (Dickey \& Lockman
1990).


The fitted parameters of the absorbed power law are rest-frame column
density N$_H^{int}=2.8^{+3.0}_{-1.4} \times 10^{22}$ \nh~ and photon
index $\Gamma=1.75^{+1.14}_{-0.20}$ (uncertainties are 90\% confidence
for one parameter of interest, $\Delta\chi^2$=2.7).  This is the
spectrum of the hard X-ray point source coincident with the radio core
in Figure 1. The slope is consistent within 1$\sigma$ with the average
value measured with \asca~for other Weak Line Radio Galaxies (WLRGs),
$\langle \Gamma_{2-10~keV} \rangle =1.49$ and dispersion
$\sigma_{\Gamma}=0.30$ (Sambruna et al. 1999). The observed fluxes are
F$_{0.2-2~keV} \sim 9 \times 10^{-15}$ and F$_{2-10~keV} \sim 1.8
\times 10^{-13}$ \flux. The intrinsic (absorption-corrected)
luminosity is L$_{2-10~keV} \sim 1.2 \times 10^{42}$ \lum, at the
high-end of the distribution for WLRGs and LINERs.

The data require a thermal component at soft energies, with fitted
temperature $kT=1.05^{+0.32}_{-0.14}$ keV, consistent with the value
measured for other radio sources with \rosat~and \asca~ (Worrall \&
Birkinshaw 1994; Sambruna et al. 1999). The abundance was fixed to the
best-fit value, 0.1 solar. The observed fluxes are F$_{0.2-2~keV}
\sim 3 \times 10^{-14}$ and F$_{2-10~keV} \sim 5 \times 10^{-15}$
\flux. A possible origin of the thermal component is the halo of the
host galaxy. Indeed, the intrinsic luminosity is L$_{0.2-2~keV} \sim 2
\times 10^{41}$ \lum, consistent with the values measured for
elliptical galaxies. Alternatively, the thermal component could be due
to the cooling flow.  


\section{Discussion and Conclusions}

Hydra~A is the second powerful radio galaxy observed with \chandra,
and the second instance where a nuclear X-ray point source is
detected. A luminous AGN was also discovered in an ACIS-S 20 ks image
of the FR~II radio galaxy 3C~295 (Harris et al. 2000), with $\Gamma
\sim 1.8$ and $L_{0.2-10~keV} \sim 7 \times 10^{43}$ \lum. As in
Hydra~A, the AGN in 3C~295 resides in a cooling flow. These results
confirm that optically-weak, narrow-emission line radio sources harbor
true AGNs as their brighter, broad-emission line Seyfert-like 
counterparts. Thanks to \chandra, a detailed study of the central
engines of these systems becomes possible for the first time.

The large X-ray column density we measure in the nucleus of Hydra~A,
N$_H^{int} \sim 3 \times 10^{22}$ \nh, implies an optical extinction
$A_V=10$, assuming galactic gas-to-dust ratios (Bohlin, Savage, \&
Drake 1978). This means that the optical and UV continuum from the AGN
are strongly (factor $\gtrsim 300$) suppressed, accounting for why the
AGN was not previously detected at these wavelengths. The weak UV
continuum detected in archival IUE spectra (Hansen et al. 1995 and our
own inspection of an unpublished spectrum) cannot be due to the AGN;
most likely candidates are the starburst and/or the cooling flow.
Future {\it HST} observations should detect only an extended thermal
component to the UV light.
 
We now turn to the issue of whether the LLAGN is capable of
photoionizing the emission-line nebula identified with the nucleus
(Baum et al. 1989; Heckman et al.  1989). To make this assessment we
first evaluate the total reddening to the nuclear emission-line
source, $E_{\rm B-V}$. Since the nuclear spectrum resembles that of
LINERs, we assume that the intrinsic Balmer decrement is
H$\alpha$/H$\beta=3.1$ (e.g., Ho, Filippenko, \& Sargent 1997b), leading
to an estimate of $E_{\rm B-V}\approx 0.33$\footnote{This is
considerably higher than $E_{\rm B-V}=0.15$ given by Hansen et
al. (1995) based on the Ly$\alpha$/H$\beta$ ratio. Hansen et al. have
probably underestimated the reddening because the Ly$\alpha$ flux was
measured in the large IUE aperture which very likely includes a
contribution from the circumnuclear starburst region.}.  With this
value of the reddening and the H$\beta$ flux measured by Hansen et
al. (1995) we obtain the rate of emission of H$\beta$ photons from the
nebula as $Q_{\rm H\beta}=4.1\times 10^{51}$~s$^{-1}$. If the H$\beta$
emission is the result of case~B recombination, then atomic physics
implies $Q_{\rm H\beta}=0.12\; Q_{\rm ion}$ (Osterbrock 1989), where
$Q_{\rm ion}$ is the ionizing photon rate from the active nucleus.
Thus, the observed H$\beta$ flux requires that $Q_{\rm ion}=(3.4\times
10^{52}/f_{\rm c})$~s$^{-1}$, where $f_{\rm c}$ is the covering
fraction of the source by the nebula. In contrast, the observed
ionizing photon rate, assuming that the X-ray power-law spectrum
extends all the way through the UV band, is $Q_{\rm ion}=1.1\times
10^{52}$~s$^{-1}$, which falls short of the required rate by a
considerable factor, even if $f_{\rm c}=1$! In fact, if we model the
spectral energy distribution (SED) of Hydra A after those observed in
LINERs and other LLAGN, some of which are radio-loud (Ho 1999b), the
problem persists.

In order to balance the photon budget of the nebula, we must either
postulate that the ionizing spectrum includes a UV bump or invoke and
additional power source. If we assume there is a UV bump, we can
parameterize it following Mathews \& Ferland (1987) and normalize the
SED according to the \chandra~ X-ray spectrum. We then find an
ionizing photon output rate of $Q_{\rm ion}=6.4\times
10^{53}$~s$^{-1}$, which is more than enough to power the emission
lines. 
On the other hand, one or more other power sources may contribute to
the ionization of the nebula.  X-rays from shocks in the cooling flow
are plausible (e.g., Heckman et al. 1989). Another possibility is
mechanical interaction of the emission-line gas with the radio jets,
which is particularly relevant since the images of Hansen et
al. (1995) show superpositions of radio and emission-line knots.

To get a handle on the nature of the accretion flow, we compared the
X-ray luminosity to the limiting Eddington luminosity through a rough
estimate of the central black hole mass in Hydra~A, $M_{BH}$.  The
apparent $V$ magnitude of the host galaxy in Hydra~A, $m_V=13.7$
(Sandage 1973), and Figure 8 of Magorrian et al. (1998) imply $M_{BH}
\sim 4 \times 10^9 M_{\odot}$, consistent with the values dynamically
measured in other giant ellipticals (e.g., Ford et al. 1994).  The
Eddington luminosity is thus $L_{Edd} \sim 5 \times 10^{47}$
\lum. Assuming $L_X=0.1 L_{total}$, the luminosity relative to
Eddington is $L_{total}/L_{Edd} \sim 2 \times 10^{-5}$. This places
Hydra~A in the regime of an Advection-Dominated Accretion Flow (ADAF;
e.g., Narayan, Mahadevan, \& Quataert 1998), similar to other WLRGs
(Sambruna et al. 1999). A powerful diagnostic of the accretion flow
structure in Hydra~A will be afforded by future higher-sensitivity
X-ray observations, such as delivered by {\it XMM}. The EPIC spectrum
will allow us to study in more detail the nuclear X-ray properties, in
particular whether an Fe line is present at 6--7 keV. The line energy
and profile could allow us to discriminate between a standard
Seyfert-like disk and an ADAF (e.g., Sambruna \& Eracleous 1999).


Finally, we note that the nuclear X-ray absorption in Hydra~A is much
larger (factor 10) than the optical/UV extinction to the emission-line
nebula (see above), and consistent with the column measured in the
radio toward the core (Taylor 1996). This suggests that the X-ray
absorber lies further in than the emission-line regions, close to the
central black hole. A possible candidate is the molecular torus on
parsec scales postulated by unification models (Urry \& Padovani
1995). Indeed, the radio and optical observations indicate an edge-on
geometry for Hydra~A (Taylor et al. 1990; Baum et al. 1989), such that
the torus intercepts the line of sight to the nucleus.  Our results
thus support current unification models for radio-loud sources and in
particular the presence of an obscuring torus in FR~I radio
galaxies. Clearly, large unbiased samples are needed to reach firmer
conclusions, which we anticipate from \chandra~in the next few years.




\acknowledgements

RMS acknowledges support from NASA contract NAS--38252. We are
grateful to Gordon Garmire and the ACIS team for making these
observations possible. We thank Joe Pesce for help with Figure 2, Niel
Brandt for the \verb+ASMOOTH+ routine, and Pat Broos and Scott Koch
for assistance with data retrieving and for the \verb+TARA+ software.
Finally, we are grateful to the referee, Yuichi Terashima, for his
prompt and thoughtful comments and suggestions.

\newpage

\noindent{\bf Figure Captions}

\begin{itemize}
 
\item\noindent Figure 1: \chandra~ACIS-S image in 0.2--10 keV of the
central regions of Hydra~A, in a 20 ks calibration exposure. Radio VLA
contours at 6 cm are overlayed (Taylor et al. 1990).  The ACIS-S image
was adaptively smoothed using a Gaussian function with a kernel of
3.5\arcsec, and shifted by 3.6\arcsec~in declination to align the
X-ray and radio core. X-ray emission from the compact radio core is
apparent, embedded in a diffuse halo, while the jets and lobes occupy
regions deficient of X-rays. 

\item\noindent Figure 2: Contours of the inner region of Hydra~A at 4
keV. North is up and East is to the left.  The intensity scale is
logarithmic with contour interval 0.4 or a factor 2.5 in
intensity. The nuclear point source is apparent. A faint extended
structure is also present at $\sim$ 3\arcsec~ N-W of the nucleus,
roughly at the position of the star formation region previously
detected in optical images (McNamara 1995; Melnick et al. 1997).

\item\noindent Figure 3: The \chandra~spectrum of the nuclear region
in Hydra~A, extracted in a region of radius 1.5\arcsec. The top panel
shows the data convolved with the best-fit model, a highly absorbed
power law at energies $\gtrsim$ 2 keV and a soft thermal component
(dotted lines). The bottom panel are the residuals in the form of
ratio of the data to the model. Crosses are ACIS-S data, asterisks are
ACIS-I data.

\end{itemize}


\newpage 

\begin{references}

\reference {} Baum, S. A., Heckman, T. M., Bridle, A. H., van Breugel,
W. J. M., \& Miley, G. K. 1988, ApJS, 68, 833 

\reference {} Bohlin, R. C., Savage, B. D., \& Drake, J. F. 1978, ApJ,
224, 132 



\reference {} David, L. P., Arnaud, K., Forman, W., \& Jones, C. 1990,
ApJ, 356, 32 

\reference {} Dickey, J. M. \& Lockman, F. J. 1990, ARAA, 28, 215 



\reference {} Fabbiano, G. 1996, in The Physics of LINERs in View of
Recent Observations, ASP Conference Series No. 103 eds. M. Eracleous,
et al. (San Francisco: ASP), p.56

\reference {} Fabian, A. C. \& Rees, M. J. 1995, MNRAS, 277, L55 



\reference {} Ford, H. C. et al. 1994, ApJ, 435, L27 




\reference {} Hansen, L., J{\o}ergsen, H. E., \& N{\o}rgaard-Nielsen,
H. U. 1995, A\&A, 297, 13 

\reference {} Harris, D. et al. 2000, ApJ, in press (astro-ph/9911381) 

\reference {} Heckman, T. M. 1986, PASP, 98, 159 

\reference {} Heckman, T. M., Baum, S. A., van Breugel, W. J. M., \&
McCarthy, P. 1989, ApJ, 338, 48 

\reference {} Ho, L. C. 1999a, ApJ, 510, 631 

\reference {} Ho, L. C. 1999b, ApJ, 516, 672 

\reference {} 
Ho, L. C., Filippenko, A. V., \& Sargent, W. L. W. 1997a, ApJ, 487, 568

\reference {} 
Ho, L. C., Filippenko, A. V., \& Sargent, W. L. W. 1997b, ApJS, 112, 315

\reference {} Ikebe, Y. et al. 1997, ApJ, 481, 660 

\reference {} Jerius, D. (ed.) 2000: ``XRCF Phase 1 Testing: Analysis
Results'', Technical Report, Smithsonian Astrophysical Observatory, in
preparation
(http://hea-www.harvard.edu/MST/simul/xrcf/report/index.html)

\reference {} Kormendy, J. \& Richstone, D. 1995, ARAA, 33, 581 


\reference {} Makishima, K. et al. 1994, PASJ, 46, L77 

\reference {} Magorrian, J. et al. 1998, AJ, 115, 2285 

\reference {} Mathews, W. G. \& Ferland, G. J. 1987, ApJ, 323, 456

\reference {} McNamara, B. R. 1995, ApJ, 443, 77 

\reference {} McNamara, B. R. et al. 2000, ApJL, subm. (astro-ph/0001402) 

\reference {} Melnick, J., Gopal-Krishna, \& Terlevich, R. 1997, A\&A,
318, 337 

\reference {} Narayan, R., Mahadevan, R., \& Quataert, E. 1998, in The
Theory of Black Hole Accretion Disks, eds. M.A.Abramowicz,
G.Bjornsson, \&J.E.Pringle (Cambridge University Press)
(astro-ph/9803141)


\reference {} Osterbrock, D. E. 1989, Astrophysics of Gaseous Nebulae
and Active Galactic Nuclei (Mill Valley: University Science Books),
326

\reference {} Ptak, A., Serlemitsos, P., Yaqoob, T., \& Mushotzky,
R. 1999, ApJS, 120, 179 



\reference {} Sadler, E. M., Jenkins, C. R., \& Kotanyi, C. G. 1989,
MNRAS, 240, 591

\reference {} Sambruna, R. M. \& Eracleous, M. 1999, in Proc. of the
X-ray Astronomy 1999 Meeting held in Bologna, Italy, 1999 September,
Astrophysical Letters and Communications, in press (astro-ph/9911503)

\reference {} Sambruna, R. M., Eracleous, M., \& Mushotzky,
R. F. 1999, ApJ, 526, 60 

\reference {} Sandage, A. 1973, ApJ, 183, 731 


\reference {} Tadhunter, C. N., Morganti, R., Robinson, A., Dickson,
R., Villar-Martin, M., \& Fosbury, R. A. E. 1998, MNRAS, 298, 1035 

\reference {} Taylor, G. B., Perley, R. A., Inoue, M., Kato, T.,
Tabara, H., \& Aizu, K. 1990, ApJ, 360, 41 

\reference {} Taylor, G. B. 1996, ApJ, 470, 394 

\reference {} Terashima, Y., Kunieda, H., \& Misaki, K. 1999, PASJ,
51, 277

\reference {} Urry, C. M. \& Padovani, P. 1995, PASP, 107, 803 

\reference {} Worrall, D. M. \& Birkinshaw, M. 1994, ApJ, 427, 134 

\end{references}
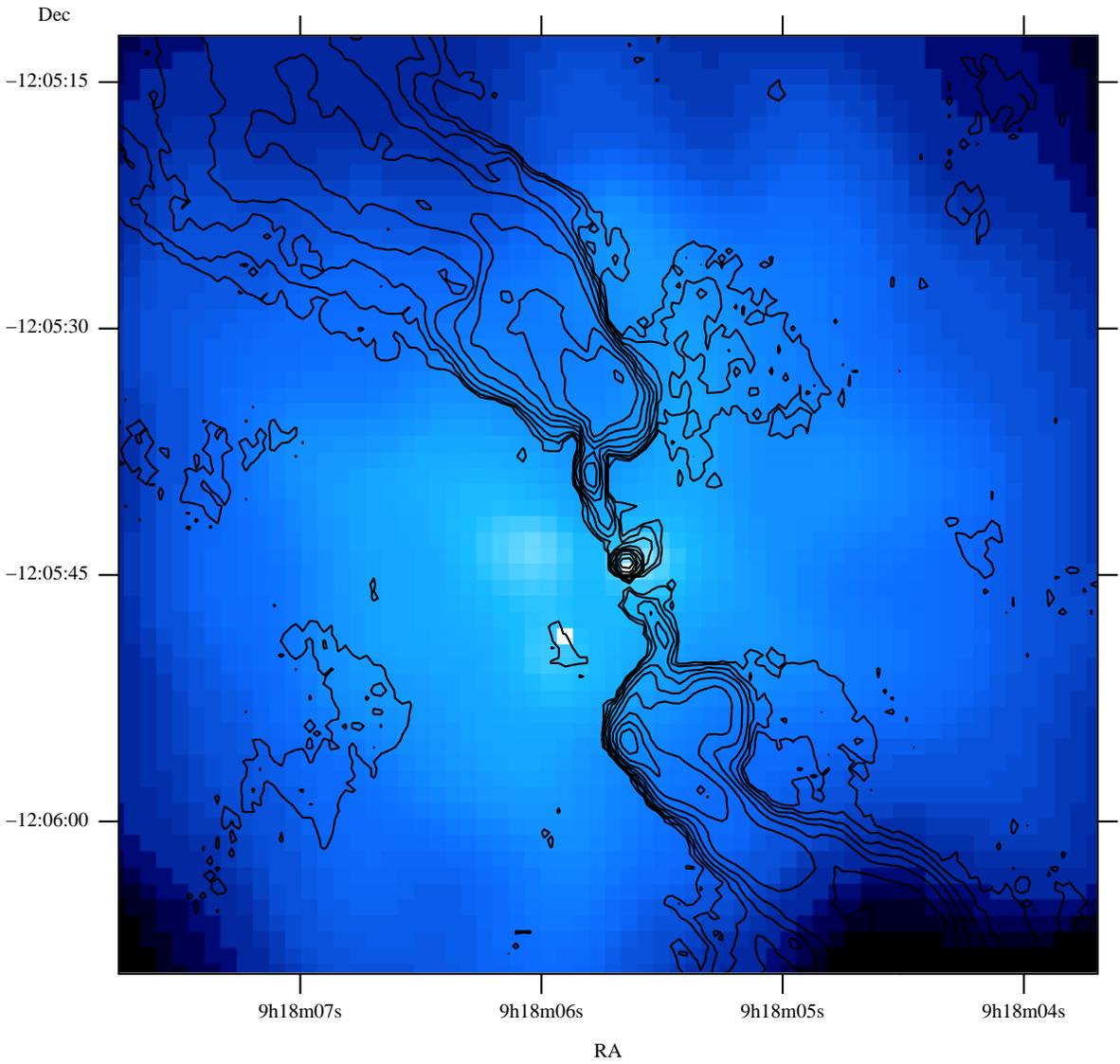
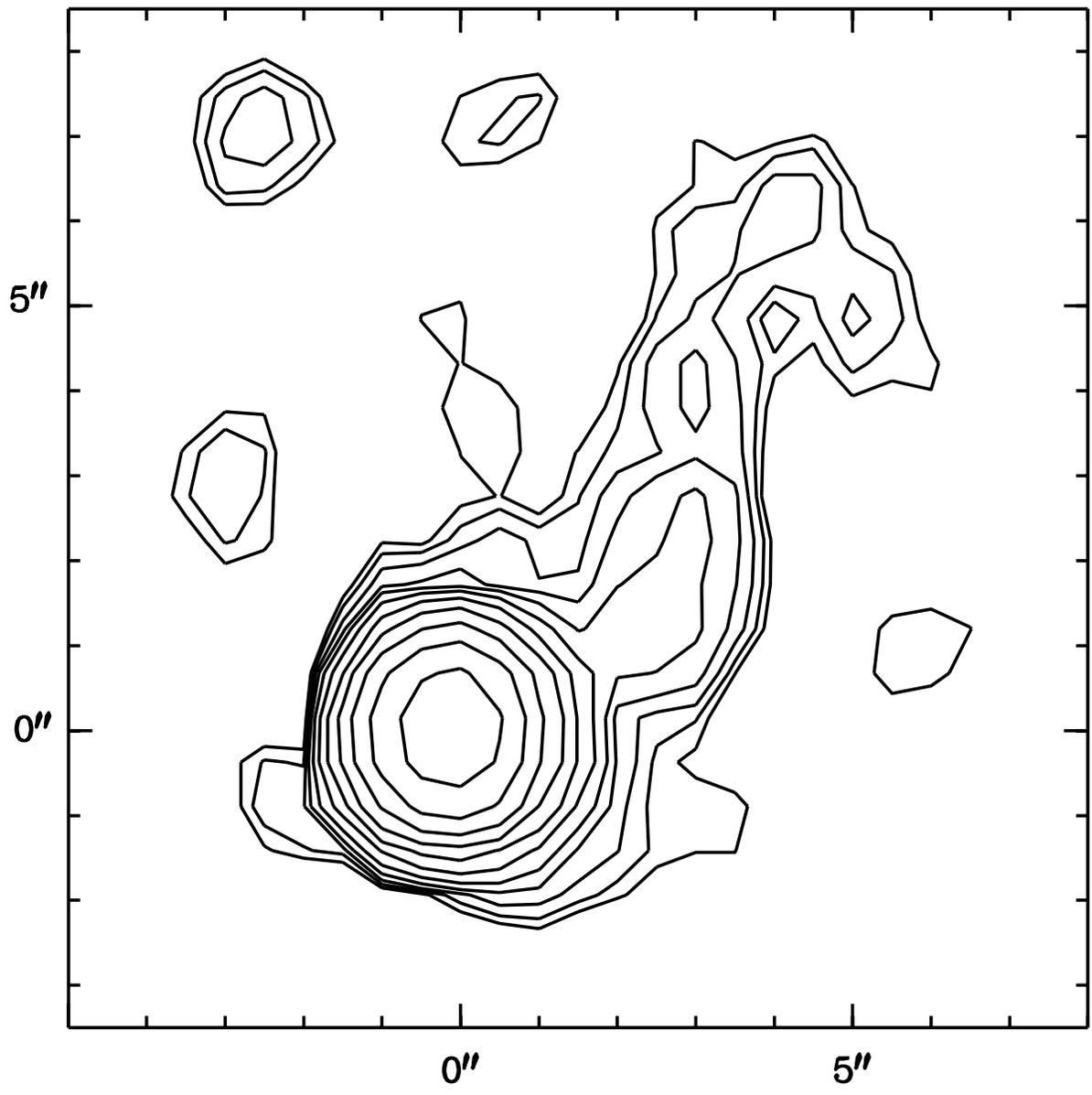
\end{document}